# Adaptive Fixed Priority End-To-End Imprecise Scheduling In Distributed Real Time Systems

Dr. W.A.EL-Haweet

Alexandria University
Faculty of Engineering
Computer science &
Automatic control Dept.

W_Elhaweet@yahoo.com

Eng. Islam.M.Elgedawy

Alexandria University
Faculty of Engineering
Computer science &
Automatic control Dept.

Islam612@yahoo.com

Prof.Dr. Ibrahim Abd EL-Salam

Alexandria University
Faculty of Engineering
Computer science &
Automatic control Dept.

i_abdelsalam@yahoo.com

## Abstract

In end-to-end distributed real time systems, a task may be executed sequentially on different processors. The end-to-end task response time must not exceed the end-to-end task deadline to consider the task a schedulable task. In transient over load periods, deadlines may be missed or processors may saturate. The imprecise computation technique is a way to overcome the mentioned problems by trading off precision and timeliness. We developed an imprecise integrated framework for scheduling fixed priority end-to-end tasks in distributed real time systems by extending an existing integrated framework for the same problem. We devised a new priority assignment scheme called global mandatory relevance scheme to meet the concept of imprecise computation. We devised an algorithm for processor utilization adjustment, this algorithm decreases the processor load when the processor utilization is greater than one. Also we extended the schedulability analysis algorithms presented in the old framework to allow adaptive priority assignment and to meet imprecise computation concept. Simulation results showed that our new framework is more dependable and predictable than the existing framework over transient overload periods.

## 1. Introduction

In [1] an integrated framework for scheduling fixed priority end-to-end tasks in distributed real time systems is proposed. This framework solves the following three related problems: Priority Assignment, Execution Synchronization and Schedulability Analysis which is based on the busy period analysis proposed in [2]. This framework fails if the task's worst-case response time analysis exceeds its end-to-end deadline or if the processor (resource) utilization exceeds one. The imprecise computation technique that is presented in [3] is used to overcome the mentioned problems. In [4] an extended imprecise computation model is proposed. We developed an adaptive imprecise integrated framework for scheduling fixed priority end-to-end tasks in distributed real time systems by combining work in [1,4] to overcome problems appeared in [1]. We devised a new priority assignment scheme called global mandatory relevance scheme and devised an algorithm for processor utilization adjustment. Also we extended the schedulability analysis algorithms' functionality presented in [1]. To overcome the first problem, we try to reduce the response time of the task to meet its deadline. The task in distributed real time systems can be viewed as a chain of subtasks that work on different processors. The output quality is a function of task's processing time and the input quality. So the extended imprecise computation model presented in [4] is used to add the advantage of vanquishing the input error, to our work. In the extended imprecise computation model each task $T_i$ is decomposed into two parts mandatory part $M_i$ and optional parts $O_i$ that can be extended according to $h_i$ and $k_i$, the mandatory scaling factor and optional scaling factor, respectively. We try to solve the processor utilization adjustment problem by reducing the execution time of subtasks that execute on the failed processor. Our imprecise integrated framework satisfies the required issues mentioned in [5]. The rest of this paper is organized as follows: In section 2, we state the used system model. In section 3, we introduce the new priority assignment schemes. In section 4, we present the processor utilization adjustment algorithm. In section 5, we present the extended schedulability analysis algorithms, discussing the imprecise schedulability analysis algorithm. In section 6, we present the simulation experiments and finally we give conclusion in section 7.

## 2. System Model

Our system model extends the system model in [1] using the extended imprecise computation model in [4] and adds new derived parameters MR (mandatory relevance), PI (priority index), VD (virtual deadline). Model assumptions and other parameters are the same as in [1,4]. We used adaptive fixed priority assignment that increases system schedulability. Subtasks inherit their parent priority (composite task).

## 3.1 Global Mandatory Relevance Priority Assignment Scheme

It was proved in [1] that every sub-problem of the priority assignment problem for the case of any synchronization protocol and schedulability analysis algorithm is NP-hard. So there are only heuristic solutions. The heuristic schemes, which are proposed in [1], are not appropriate to be applied on the new framework because it gives illogical results as indicated in [7]. Summarizing issues to be considered in the new priority assignment scheme:
(1) A task is decomposed in two parts mandatory and optional. The mandatory part has much higher priority than the optional part. (2) The mandatory part's relevance is the mandatory execution time compared to the whole task execution time (3) Follow a deadline-based scheme. The new priority assignment scheme compromises among the above three issues, giving the highest priority to the task that has the highest mandatory relevance and shortest deadline.

### Concept of the new scheme

The composite tasks are going to be sorted from the highest to the lowest priority according to certain criterion. Hence, the sort position of the parent composite task represents the priority of its subtasks. Now the new factors that are used to determine the task's priority will be introduced.

**Mandatory Relevance (MR):** Is the ratio between the task's mandatory part and the whole task's execution time

$$Mandatory\,\text{Re}\,levance(T_i) = \frac{\sum_{j=1}^{j=n} M_{i,j}}{\sum_{j=1}^{j=n} M_{i,j} + O_{i,j}}$$

Where $M_{i,j}$ and $O_{i,j}$ are the mandatory and optional parts of $T_i$ subtasks and n is the number of the subtasks.

**Virtual Deadline (VD):** Is a variable that is initiated by the value of the relative end-to-end deadline of the composite task $T_i$. The value of Virtual Deadline can be decreased during the computation according to the system conditions to increase task priority. Its value never affects the relative end-to-end deadline and schedulability analysis does not use it in any deadline comparison. It is just used in task's priority promotion.

**Priority Index (PI):** Is the criterion used in tasks sorting. Priority Index is computed from the following equation:

$$PI(T_i) = \frac{Mandatory\,\text{Re}\,levance(T_i)}{VirtualDeadline(T_i)}$$

**Scheme Implementation**

The composite tasks are going to be sorted according to their priority index. But one can get many tasks having the same priority indices. In this case the algorithm gives the composite task with the shorter relative end-to-end deadline the higher priority, because it follows the deadline-based schemes logic, as well equal priority indices and equal relative end-to-end deadline can exist. In this case, the scheme gives the composite task with the shorter mandatory part the higher priority, because this means they have the same mandatory relevance. Hence the shorter mandatory part task is the shorter execution time task, the algorithm gives it the higher priority to increase throughput following the concept of shortest job first. After sorting the composite tasks, subtasks inherit their parent priority. This scheme is a global priority scheme because the composite task's priority is determined according to interaction with all the system's tasks giving an accurate status about the task's priority. If global information is missing e.g. on-line scheduling, a local priority scheme is going to be used in which subtasks have different priority than their parent, for that reason another variation of this scheme is presented.

## 3.2 Local Mandatory Relevance Priority Assignment Scheme

The same concept of the global scheme can be applied on every processor's tasks, not on all tasks in the system. By calculating the mandatory relevance of the subtask as the ratio of its mandatory part compared to its execution time. Hence, the priority index is given by

$$PI(T_{i,j}) = \frac{M_{i,j}}{(M_{i,j} + O_{i,j}) * VD(T_i)}$$

## 4. Processor Utilization Adjustment

The framework proposed in [1] fails if the processor utilization exceeds 1. One has an opportunity to make the system schedulable by using the imprecise computation technique. When a processor fails, one tries to reduce the execution time of its subtasks. This is done by reducing the execution time of optional parts of its subtasks then try to eliminate the input error effect due to optional part reduction by extending both mandatory and optional parts of the successor subtask according to the task error functions [4]. One starts reducing the optional part of the least priority subtask until processor utilization be less than 1 or continue taking from optional parts of the higher priority subtasks until we reach the highest priority subtask. If all optional parts of the subtasks are reduced and the processor still fails, the system will fail. This means that the system is not schedulable with the current configuration. When one tries to eliminate the effect of input error the following may happen:

- The extensions that are made for successor subtasks do not cause any other processors failure. In this case, what is lost in one processor is gained on another one making final error equals zero. This makes better utilization of the system resources.

- One can not eliminate the input error because the reduced subtask is the last subtask in the chain and no successor for it. In this case, one continues scheduling the system knowing the final result will be with acceptable error.

- The extensions that are made for successor subtasks cause other processors failure. So one repeats the utilization adjustment action for the failed processor that has the highest priority and the least subscript subtask executes on it. This choice is made to give a chance for the higher priority tasks to survive and limit the execution time of lowest priority tasks.

One repeats the above until system reports failure or success.

If we have N composite tasks and P processor the complexity of the algorithm is O(NP) [7].

## 5. Extended Schedulability Analysis Algorithms

As mentioned before, the old integrated framework also fails when the worst case response time of the composite task exceeds its relative end-to-end deadline. Concerning the trade off between precision and timeliness we are going to reduce the composite task response time using imprecise computation to make the composite task meets its deadline. The response time of the composite task consists of two parts:

- **Execution Time:** The sum of all its subtasks 'execution time (mandatory + optional).

- **Blocking Time:** The sum of all time periods that the task is preempted by higher subtasks.

Hence, to reduce the response time one can reduce the execution time, blocking time or both.

### 5.1 Response Time Reduction:

**Execution Time Reduction phase**

In this phase we reduce the optional parts of the failed composite task to make it schedulable with the least possible final error, taking into consideration the extensions due to optional reduction. In [4] near optimal algorithms are used to reassign task's optional parts obtaining the least possible final error. We use these algorithms to solve our problem. Five algorithms are proposed Dist_M, Dist_M+ or Dist_M+Iterative which are used when the mandatory extension is bigger than the optional extension and Dist_O or Dist_O+ which are used when the optional extension is bigger than the mandatory extension [4]. It is important to notice that deadline that is passed to the algorithms must equal to the difference between task's deadline and the sum of blocking time of its subtasks, to preserve the blocking time effect on the task response time after optional parts reductions.

After applying above algorithms one can have the following cases:

- The algorithms failed to reduce the optional parts to make the composite task meet its deadline. In this case one assigns mandatory parts to all subtasks taking into consideration the mandatory extensions due to input error, and starts working in phase two (blocking time reduction or task promotion).
- The algorithms succeeded to reduce optional parts with minimal possible error and no processor failed due to the execution time reassignments.
- The algorithms succeeded to reduce the optional parts but due to execution time reassignment some processors failed. In this case, one assigns mandatory parts to all subtasks taking into consideration the mandatory extensions, and tries to give the last subtask as optional as it can take as long as it does not violate its deadline then let the utilization adjustment algorithm handles the case.

One assigns mandatory parts to all subtasks except the last one just to make the job easy for the algorithms in the next iteration and if optional parts will be reduced, the reduction would be from the last subtask only.

After the execution phase is finished, one continues with the next phase, to see the effect of execution time reduction over the tasks' priority. Even if the execution time phase is failed, one takes the information about the amount of time needed to make the composite task meets its deadline (we refer to this amount as shortage) and it will be used in the next phase.

## 5.2 Blocking Time Reduction

When the execution time reduction fails, this means that the task has only mandatory parts and still missing its deadline. In this case the blocking time must be reduced. By reducing the execution time of higher priority tasks we can make the lower priority task schedulable.

So it is needed to reduce the optional parts of higher priority tasks. This is not an easy problem to solve, because it demands to see the effect of this reduction, taking into consideration the extension in execution time of the successors, which might cause many problems. For example processors may fail (utilization >1), other subtasks on other processors may get their blocking time increased which may make other composite tasks to miss their deadline.

The concept of using the integrated framework will be violated if one tries to deal with all of these problems in the same time. The exit from this trap is to promote the failed task (increase its priority). Hence, the number of tasks with higher priority will be reduced giving a chance for the task to meet its deadline.

Task's priority promotion will increase system schedulability that instead the schedulability analysis algorithm stuck with one priority assignment as in [1], another priority assignment can be found to make it schedulable.

Now how can one promote the task? This depends on the task status and its current configuration.
The following cases arise:

- The execution time reduction phase is succeeded and the optional part is decreased so the mandatory relevance is increased. The task's priority is recalculated. The task may be promoted and the system is schedulable.

- The execution time reduction fails, two actions are made to promote the task:
  1- Task mandatory relevance becomes 1
  2- The task's virtual deadline is decreased by the shortage reported from the execution time reduction phase as long as the virtual deadline value >0.
- The execution time reduction fails, the task mandatory relevance = 1 and (VD – shortage) < 0. This means that this task must be the highest priority task in the system. In this case the following should be done:
  1- Make task's PI = 1.
  2- Make its VD = relative deadline of the task
  3- Mark the task as depleted task

The depleted task can not be reduced any more, also can not be promoted any more because the above configuration makes it the highest priority task in the system.

In task's priority recalculation, if there are more than one depleted task in the system they are arranged according to the rules indicated in the global mandatory scheme.

## 5.3 Imprecise Extended Schedulability Analysis Algorithm

The old schedulability analysis algorithm's function is to determine whether the system is schedulable or not [1]. But when we integrate the concept of imprecise computation with it, new functions are added to the schedulability analysis algorithm such as the reduction of the response time, the processor utilization adjustment and priority assignment adaptation.

The imprecise extended schedulability analysis algorithm tries to get the system configuration that makes it schedulable with the available resources. The imprecise extended schedulability analysis algorithm is an iterative algorithm, each iteration has the following steps:

- According to the current system configuration, it searches for failed processors (utilization >1) and fixes this problem by calling the Utilization Adjustment Algorithm. The result of this call will be correcting the failed processors or reporting system failure.

- When the system has no failed processors, one can apply the old framework [1] to solve end-to-end problem using our priority assignment scheme.

- If all the system tasks are depleted and missing deadlines still exist, the system reports failure because the system's resources are not enough.

- If there are tasks missed their deadlines and not all the system tasks are depleted. One starts with highest priority failed task to reduce its response time. Since higher priority tasks affect on the response time of lower priority tasks (increases the blocking time), so when higher priority task is fixed, lower priority tasks could be fixed automatically. Another reason for choosing higher priority to be reduced first is to minimize the number of reduced tasks. If a higher priority task is fixed first, the lower one can be fixed automatically without any response time reduction, but if we fix lower priority task first we will have to fix the higher priority task too.

- If the highest priority failed task is a depleted task then the system fails.

- After correcting the highest priority failed task, one takes another iteration to get the final status of the system, if success one stops the algorithm else one repeats the work for the next highest priority tasks that missed its deadline.

In [1] the resource contention problem is solved using the old framework of end-to-end problem by applying a simple mapping algorithm. One can apply our framework to solve the resource contention problem using the same mapping algorithm.

### 5.4 Algorithm Complexity

**Lemma**

If the system contain N composite tasks, and each composite task has at most M subtasks and number of the processors are P. The imprecise schedulability analysis algorithm worst case complexity will be O($N^2(M+P)$) [7].

## 6. Simulation Results:

We study the performance of the suggested imprecise integrated framework and compare it with the old framework in [1] to show how the dependability of the system is increased. Five simulation experiments are performed using the transient overload work models presented in [6]. For more details about workload generation, simulation experiments, simulator validation and verification and algorithms pseudo code refer to [7].

**Performance Criteria**

We used the performance criteria in [1,4] to validate our work. Work in [1] used the schedulability index and failure rate, work in [4] used the average final error and we added the worst final error to show the maximum error occurred in the system indicating the trade off between imprecision and timeliness. All the experiments compare the reaction of the two frameworks under the experiment conditions.

**Experiment 1:** It starts with certain deadline and decrease it gradually.

**Experiment 2:** The **Balanced Overload** work model is used.

**Experiment 3:** The **Unbalanced Overload** work model is used. We repeat it twice one for optional parts extension and other for mandatory part extensions.

**Experiment 4:** The **Task-Set-Increase Overload** work model is used.

**Experiment 5:** The **Frequency-Increase Overload** work model is used.

### Comments on results

**Experiment 1:** The results of experiment 1 are indicated in Table 1. From the results we see that the failure rate and schedulability index are noticeably reduced using the imprecise framework, hence the system dependability is increased. The failure rate equals zero for the imprecise framework until we reduced the deadline by 70% the failure rate starts to increase which mean that the deadline in this stage is not enough for scheduling only mandatory parts of the tasks. The failure rate approximately reaches 1 when we reduced 90% of the deadline. So we succeed to make the system schedulable when the normal framework fails. We notice that the worst-case final error equals 1 when percentage exceed 70% means there are tasks that loose all its final optional part. Also we notice that the average final error exceeds as the percentage increases which means many optional parts are decreased to meet the deadline

**Experiment 2:** The results of experiment 2 are indicated in Table 2. From the results we see that the failure rate and schedulability index are reduced using the imprecise framework, hence the system dependability is increased. Failure rate equals zero in the imprecise case because even when the utilization of a processor exceeds 1, the optional part is discarded and the system still schedulable. We notice that the worst-case final error equals 1 when the scale exceed 3 means that there are tasks loose all its final optional part. Also we notice that the average final error exceeds as the scale increases which means many optional parts are decreased to make system schedulable.

**Experiment 3:** The results of experiments 3a and 3b are listed in Tables 3,4 consequently, when we scaling the optional part (experiment 3a) we see that the failure rate and schedulability index are reduced using the imprecise framework. Failure rate equals zero in the imprecise case because even when the utilization of a processor exceeds 1, the optional part is discarded and the system still schedulable. But when we scale the mandatory part (experiment 3b) we see the schedulability index of the imprecise framework is bigger or equals to the old framework. This happens because when we scale mandatory part we increase the task priority, hence increasing the blocking time of the lower priority tasks that increases the worst-case response time. Also we notice the failure rate starts to increase because when utilization exceeds 1 and no more optional time to sacrifice with, the system fails. Also we notice that the average final error exceeds as the scale increases which means many optional parts are decreased to make system schedulable.

**Experiment 4:** The results of experiment 4 are listed in Table 5, we notice that the failure rate is noticeably decreased but the schedulability index is increased because we add high priority tasks to the system hence increasing the worst-case response time. Also we notice that the average final error and worst-case final error exceeds as the scale increases which means many optional parts are decreased to make system schedulable.

**Experiment 5:** The results of experiment 5 are listed in Table 6, we notice that the failure rate is noticeably decreased and reached zero but the schedulability index value is increased then decreased. This is because schedulability index is the ratio between the response time and the period. Imprecise framework decreases the response time but we also decrease the period. So schedulability index is increased at first because response time is not decreased heavily but the period decreased giving higher schedulability index but later the amount of response time reduction is increased so the amount of period reduction giving slight difference from schedulability index in the beginning of the experiment.

### 7. CONCLUSION

We devised the following: new priority assignment schemes global and local mandatory relevance schemes, a new algorithm for processor's utilization adjustment, a new algorithm for response time reduction based on Dist_M (variations) and Dist_O (variations) algorithms and minimize the blocking time by task's priority promotion. We extend the old schedulability analysis algorithms' functionality. From simulation results we can say that we made an imprecise adaptive integrated framework for end-to-end distributed real time systems that increases system's dependability and ensures graceful degradation of the system which is not maintained in framework proposed in [1].

**Table 1**

| | Reduction Percent | 0% | 10% | 20% | 30% | 40% | 50% | 60% | 70% | 80% | 90% |
|---|---|---|---|---|---|---|---|---|---|---|---|
| **Normal Frame Work** | Failure Rate | 0.023 | 0.055 | 0.111 | 0.218 | 0.432 | 0.713 | 0.941 | 0.997 | 1.000 | 1.000 |
| | Schedulability Index | 2.948 | 2.888 | 2.805 | 2.678 | 2.468 | 2.191 | 1.826 | 1.361 | **UNS** | **UNS** |
| **Imprecise Frame Work** | Failure Rate | 0.000 | 0.000 | 0.000 | 0.000 | 0.000 | 0.000 | 0.000 | 0.009 | 0.531 | 0.999 |
| | Schedulability Index | 2.943 | 2.881 | 2.798 | 2.666 | 2.454 | 2.159 | 1.772 | 1.358 | 0.946 | 0.482 |
| | Average Final Error | 0.002 | 0.003 | 0.007 | 0.016 | 0.033 | 0.075 | 0.165 | 0.322 | 0.489 | 0.667 |
| | Worst Final Error | 0.020 | 0.041 | 0.082 | 0.178 | 0.352 | 0.624 | 0.889 | 0.993 | 1.000 | 1.000 |

**Table 2**

| | Scale | 0 | 1 | 2 | 3 | 4 | 5 | 6 | 7 | 8 | 9 |
|---|---|---|---|---|---|---|---|---|---|---|---|
| **Normal Frame Work** | Failure Rate | 0.023 | 0.023 | 0.976 | 1.000 | 1.000 | 1.000 | 1.000 | 1.000 | 1.000 | 1.000 |
| | Schedulability Index | 2.948 | 2.963 | 4.249 | UNS | UNS | UNS | UNS | UNS | UNS | UNS |
| **Imprecise Frame Work** | Failure Rate | 0.000 | 0.000 | 0.000 | 0.000 | 0.000 | 0.000 | 0.000 | 0.000 | 0.000 | 0.000 |
| | Schedulability Index | 2.943 | 2.956 | 3.989 | 3.945 | 3.903 | 3.860 | 3.836 | 3.759 | 3.736 | 3.697 |
| | Average Final Error | 0.002 | 0.002 | 0.143 | 0.383 | 0.509 | 0.582 | 0.628 | 0.665 | 0.691 | 0.710 |
| | Worst Final Error | 0.020 | 0.020 | 0.866 | 0.998 | 1.000 | 1.000 | 1.000 | 1.000 | 1.000 | 1.000 |

**Table 3**

| | Scale | 0 | 1 | 2 | 3 | 4 | 5 | 6 | 7 | 8 | 9 |
|---|---|---|---|---|---|---|---|---|---|---|---|
| **Normal Frame Work** | Failure Rate | 0.023 | 0.023 | 0.079 | 0.195 | 0.316 | 0.428 | 0.548 | 0.63 | 0.706 | 0.752 |
| | Schedulability Index | 2.948 | 2.948 | 3.167 | 3.314 | 3.366 | 3.417 | 3.412 | 3.431 | 3.340 | 3.335 |
| **Imprecise Frame Work** | Failure Rate | 0.000 | 0.000 | 0.000 | 0.000 | 0.000 | 0.000 | 0.000 | 0.000 | 0.000 | 0.000 |
| | Schedulability Index | 2.943 | 2.943 | 3.137 | 3.212 | 3.230 | 3.218 | 3.175 | 3.147 | 3.123 | 3.117 |
| | Average Final Error | 0.002 | 0.002 | 0.005 | 0.012 | 0.024 | 0.034 | 0.045 | 0.053 | 0.059 | 0.064 |
| | Worst Final Error | 0.020 | 0.020 | 0.057 | 0.136 | 0.256 | 0.344 | 0.446 | 0.513 | 0.578 | 0.622 |

**UNS: Unable to schedulable**

**Table 4**

| | Scale | 0 | 1 | 2 | 3 | 4 | 5 | 6 | 7 | 8 | 9 |
|---|---|---|---|---|---|---|---|---|---|---|---|
| **Normal Frame Work** | Failure Rate | 0.023 | 0.023 | 0.029 | 0.061 | 0.144 | 0.294 | 0.447 | 0.553 | 0.648 | 0.699 |
| | Schedulability Index | 2.948 | 2.948 | 3.042 | 3.173 | 3.291 | 3.367 | 3.365 | 3.388 | 3.353 | 3.319 |
| **Imprecise Frame Work** | Failure Rate | 0.000 | 0.000 | 0.000 | 0.000 | 0.004 | 0.011 | 0.029 | 0.066 | 0.128 | 0.199 |
| | Schedulability Index | 2.943 | 2.943 | 3.040 | 3.171 | 3.318 | 3.422 | 3.521 | 3.619 | 3.669 | 3.702 |
| | Average Final Error | 0.002 | 0.002 | 0.002 | 0.004 | 0.011 | 0.025 | 0.046 | 0.069 | 0.087 | 0.105 |
| | Worst Final Error | 0.020 | 0.020 | 0.021 | 0.047 | 0.097 | 0.205 | 0.327 | 0.408 | 0.478 | 0.513 |

**Table 5**

| | Scale | 0 | 1 | 2 | 3 | 4 | 5 | 6 | 7 | 8 | 9 |
|---|---|---|---|---|---|---|---|---|---|---|---|
| **Normal Frame Work** | Failure Rate | 0.023 | 0.053 | 0.220 | 0.514 | 0.692 | 0.797 | 0.858 | 0.885 | 0.912 | 0.924 |
| | Schedulability Index | 2.948 | 3.185 | 3.384 | 3.431 | 3.418 | 3.368 | 3.300 | 3.266 | 3.220 | 3.238 |
| **Imprecise Frame Work** | Failure Rate | 0.000 | 0.000 | 0.001 | 0.002 | 0.017 | 0.037 | 0.087 | 0.154 | 0.235 | 0.311 |
| | Schedulability Index | 2.943 | 3.191 | 3.468 | 3.696 | 3.850 | 3.927 | 3.958 | 3.961 | 3.958 | 3.973 |
| | Average Final Error | 0.002 | 0.003 | 0.016 | 0.053 | 0.104 | 0.163 | 0.217 | 0.258 | 0.292 | 0.316 |
| | Worst Final Error | 0.020 | 0.039 | 0.157 | 0.405 | 0.589 | 0.710 | 0.783 | 0.817 | 0.843 | 0.855 |

**Table 6**

| | Reduction Percent | 0% | 10% | 20% | 30% | 40% | 50% | 60% | 70% | 80% | 90% |
|---|---|---|---|---|---|---|---|---|---|---|---|
| **Normal Frame Work** | Failure Rate | 0.023 | 0.040 | 0.086 | 0.843 | 0.999 | 1.000 | 1.000 | 1.000 | 1.000 | 1.000 |
| | Schedulability Index | 2.948 | 3.339 | 4.031 | 4.664 | 5.049 | **UNS** | **UNS** | **UNS** | **UNS** | **UNS** |
| **Imprecise Frame Work** | Failure Rate | 0.000 | 0.000 | 0.000 | 0.000 | 0.000 | 0.000 | 0.000 | 0.000 | 0.000 | 0.000 |
| | Schedulability Index | 2.943 | 3.334 | 4.017 | 5.011 | 5.955 | 6.643 | 6.661 | 6.080 | 5.125 | 3.710 |
| | Average Final Error | 0.002 | 0.002 | 0.006 | 0.062 | 0.194 | 0.361 | 0.516 | 0.655 | 0.787 | 0.919 |
| | Worst Final Error | 0.020 | 0.029 | 0.075 | 0.533 | 0.945 | 1.000 | 1.000 | 1.000 | 1.000 | 1.000 |